\documentclass[aps,pre,twocolumn,showpacs,amsmath,amssymb,floatfix,amstex]{revtex4}

\usepackage[dvips]{graphicx}
\usepackage{bm,pstricks,longtable}

\input{epsf}

\def\tauq{\tau_q}
\def\tauqtyp{\tau_q^{\textrm{typ}}}
\def\dqtyp{D_q^{\textrm{typ}}}
\def\pqtyp{P_q^{\textrm{typ}}}
\def\ftyp{f^{\textrm{typ}}}
\def\calp{{\mathcal P}}
\def\calz{{\mathcal Z}}
\def\pqk{P_{q,k}}

\begin{document}

\title{Multifractal wave functions of simple quantum maps}

\author{John Martin$^1$, Ignacio Garc\'ia-Mata$^2$, Olivier Giraud$^{3,4,5}$ and  Bertrand Georgeot$^{3,4}$} 
\affiliation{
$^1$ Institut de Physique Nucl\'eaire, Atomique et de
Spectroscopie, Universit\'e de Li\`ege, B\^at.\ B15, B - 4000
Li\`ege, Belgium\\
$^2$ Departamento de F\'isica, Lab. TANDAR, Comisi\'on Nacional de Energ\'ia 
  At\'omica, Av. del Libertador 8250, 
  C1429BNP  Buenos Aires, Argentina\\
$^3$ Universit\'e de Toulouse; UPS; Laboratoire de
 Physique Th\'eorique (IRSAMC); F-31062 Toulouse, France\\
$^4$ CNRS; LPT (IRSAMC); F-31062 Toulouse, France\\
$^5$ LPTMS, CNRS and Universit\'e Paris-Sud, UMR 8626, B\^at. 100,
91405 Orsay, France}

\begin{abstract}
We study numerically multifractal properties of two models of one-dimensional quantum maps, 
a map with pseudointegrable dynamics and intermediate spectral
statistics, and a map with an Anderson-like transition recently implemented with cold atoms. Using extensive numerical simulations, we compute
the multifractal exponents of quantum wave functions and study their properties, with the help of two
different numerical methods used for classical multifractal systems
(box-counting method and wavelet method).  We compare the
results of the two methods over a wide range of values.  We show that the wave functions of
the Anderson map display a multifractal behavior similar
to eigenfunctions of the three-dimensional Anderson transition but of a weaker type. Wave functions of the intermediate map share some common properties with eigenfunctions at the Anderson transition
(two sets of multifractal exponents, with similar asymptotic behavior), but other properties
are markedly different (large linear regime for multifractal exponents even for strong
multifractality, different distributions of moments of wave functions, absence of symmetry of the exponents).  Our results thus indicate
that the intermediate map presents original properties, different from certain
characteristics of the Anderson transition derived from the nonlinear sigma model.  We also discuss the importance of finite-size effects. 
\end{abstract}
\date{July 8, 2010}

\maketitle

\section{Introduction}

Multifractal behavior has been observed in a wide
variety of physical systems, from turbulence \cite{turbulence}
to the stock market \cite{stock}
and cloud images \cite{cloud}.  It has been recognized
recently that such a behavior can also be visible in quantum wave functions 
of certain systems.  In particular, wave functions in the Anderson model of electrons in a disordered potential are multifractal
at the metal-insulator transition (see e.g. \cite{mirlin2000,cuevas,mirlinRMP08}).
Similar behaviors were seen in quantum Hall transitions \cite{huckenstein}
and in Random Matrix models such as the Power law Random Banded Matrix model
(PRBM) \cite{PRBM,kravtsov} and ultrametric random matrices \cite{ossipov}.
Such properties have also been seen in diffractive systems \cite{GG}
and pseudointegrable models,
for which there are constants of motion,
but where the dynamics takes place in surfaces more complicated than the invariant tori
of integrable systems \cite{interm}.  In all these models, this behavior
of wave functions came with a specific type of spectral statistics,
intermediate between the Wigner distribution typical of chaotic systems
and the Poisson distribution characteristic of integrable systems 
\cite{interm,girdif}.  Recently, a new model of one-dimensional
quantum ``intermediate map'' which displays
multifractal behavior was proposed \cite{giraud}, and a version with random 
phases was shown semi-rigorously to display intermediate statistics
\cite{bogomolny}.  This model is especially simple to handle numerically
and analytically, and displays different regimes of multifractality 
depending on a parameter \cite{MGG}. In parallel, new experiments allowed
recently to observe the Anderson transition with cold atoms in
an optical potential \cite{delande,delandebis} using a one-dimensional ``Anderson map'' proposed in \cite{dima}.

Although much progress has been made in the study of these
peculiar types of systems, several important questions related to
multifractality remain
unanswered.  In particular, many results were derived or
conjectured in the framework of the Anderson model, and their
applicability to other families of systems is not known.  
Also, the precise link 
between the multifractal properties of wave functions and the spectral
statistics is not elucidated.

In order to shed some light on these questions, we 
systematically investigate several properties of the wave functions
of the intermediate quantum map of \cite{giraud,bogomolny}, and compare them to 
results obtained for the Anderson map. In these one-dimensional systems, wave functions of very large vector sizes can be obtained and 
averaged over many realizations. This enables to control the errors and evaluate the reliability of standard methods used in multifractal analysis. This also allows to check and 
discuss several important conjectures put forth in the context
of Anderson transitions.  Our results also enable to study the
road to asymptotic behavior in such models, giving hints on which
quantities are more prone to finite-size effects or can be visible only with
a very large number of random realizations.  

The manuscript is organized as follows.  In Section II, we review
the known facts and conjectures about quantum multifractal
systems, which were mainly put forth in the context of the Anderson transition.
In Section III, we present the models that will be studied throughout 
the paper.  Section IV discusses the numerical methods used in order
to extract multifractal properties of the models.  Section V
presents the results of numerical simulations, allowing first to compare the 
different numerical methods of Section IV, and then to test
the conjectures and results developed in the context of the Anderson
transition to the two families of models at hand.

\section{Earlier results and conjectures}
\label{knownfacts}
We first recall some basic facts about multifractal analysis.
Localization properties of wave functions of components 
$\psi_i$, $i=1,...,N$, in a Hilbert space of dimension $N$, 
can be analyzed by means of
their moments 
\begin{equation}
\label{moments}
P_q= \sum_{i=1}^N |\psi_i|^{2q}.
\end{equation}
The asymptotic behavior of the moments \eqref{moments} for large 
 $N$ is governed by a set of multifractal exponents $\tauq$
defined by $P_q\propto N^{-\tauq}$, 
or by the associated set of multifractal dimensions $D_q=\tau_q/(q-1)$.
Equivalently, the singularity spectrum $f(\alpha)$ characterizes the fractal dimensions of
the set of points $i$ where the weights $|\psi_i|^2$ scale as
$N^{-\alpha}$. It is related to the multifractal exponents $\tauq$ by a
Legendre transform
\begin{equation}
f(\alpha)=\min_q(q\alpha -\tau_q).
\label{falpha}
\end{equation}

Compared to classical multifractal analysis,
the quantum wave functions in Hilbert space of increasing dimensions 
are considered as the same distribution at smaller and smaller scales.
This allows to define properly the multifractality of quantum wave functions, 
although at a given dimension they correspond to a finite vector.  

In many physical instances, only a single realization of the wave function
is considered. However, as soon as several realizations are considered, as is
the case in presence of disorder, the
moments \eqref{moments} are distributed according to a certain probability
distribution, and multifractal exponents depend on the way ensemble averages
are performed, and in particular on the treatment of the tail of the moments
distribution. If the tail decreases
exponentially or even algebraically with a large exponent, the two averages
should give the same answer. On the other hand, if the moments decrease
according to a power-law with a small exponent, the average
$\langle P_q \rangle$ will be dominated by rare wave
functions with
much larger moments (whose magnitude directly depends on the number of
eigenvectors considered), while the quantity 
$\pqtyp=\exp\langle \ln P_q \rangle$ will correspond to the typical value of
the moment for the bulk of the wave functions considered. To each of these
possible averaging procedures corresponds a set of multifractal exponents,
defined by 
\begin{eqnarray}
\label{scaling}
\langle P_q \rangle&\propto&N^{-\tauq}, \;\tauq=D_q(q-1),\\
\label{scalingtyp}
\pqtyp&\propto&N^{-\tauqtyp},\; \tauqtyp = \dqtyp(q-1).
\end{eqnarray}
As soon as averages over several realizations are made there can be a 
discrepancy between $\tauq$ and $\tauqtyp$. Historically this effect was
seen in the context of the Anderson transition 
\cite{MirlinEvers00,mirlinRMP08} and was very recently confirmed by
the analytical 
calculations of \cite{ludwig,MonGar10} in the same model. More specifically, 
it was shown that the distribution of the normalized moments $y_q=P_q/\pqtyp$
is asymptotically independent of $N$ and has a power-law tail
\begin{equation}
\calp(y_q)\sim\frac1{y_q^{1+x_q}}
\end{equation} 
for large $y_q$~\cite{EversMirlin00,MirlinEvers00}.
The multifractal exponents $\tauq$ and $\tauqtyp$ only coincide over an 
interval $[q_-, q_+]$ where $x_q>1$. In the case of heavy tails
$x_q<1$, the averages $\langle P_q \rangle$ and $\exp\langle \ln P_q \rangle$ yield different exponents.

This phenomenon has a counterpart in the singularity spectra  $f(\alpha)$
and $\ftyp(\alpha)$. While $\ftyp(\alpha)$ cannot take values below 0 and
terminates at points $\alpha_{\pm}$ such that $\ftyp(\alpha_{\pm})=0$, 
the singularity spectrum $f(\alpha)$ continues below zero. The two spectra
coincide over the interval $[\alpha_+,\alpha_-]$. It can be shown that 
outside the interval $[q_-, q_+]$ the set of exponents $\tauqtyp$ are given
by the linear relation $\tauqtyp=q\alpha_{+}$ for $q>q_{+}$ and 
$\tauqtyp=q\alpha_{-}$ for $q<q_{-}$ \cite{mirlinRMP08}.

In \cite{MirlinEvers00}, it was stated that the exponents $\tau_q$ and 
$\tauqtyp$ can be related through the following relation which depends
on the tail exponent of the moments distribution $x_q$:
\begin{equation}
\label{xqtq}
x_q \tauqtyp=\tau_{qx_q}.
\end{equation}
Equation \eqref{xqtq} was analytically proven for PRBM \cite{PRBM} for
integer values of $x_q$ and also in the limit of large bands
for $q>1/2$. It remains unclear to which extent Eq.~\eqref{xqtq} is valid for other types of systems. A consequence of the identity \eqref{xqtq} for $q>q_+$ is that 
$x_q=q_{+}/q$. In the regime of weak multifractality where $D_q$ is a linear
function,  the identity \eqref{xqtq} implies that $x_q=(q_+/q)^2$ for $q_-<q<q_+$ \cite{mirlinRMP08,PRBM}(see Subsection V.E. for details). 

Finally, a further relation that we wish to investigate in the present paper has been predicted based on the nonlinear sigma model and 
observed in the 3D Anderson model at criticality and several other systems.
It is a symmetry relation 
of multifractal exponents \cite{fyodorov}, which can be expressed as
\begin{equation}
\label{deltaqrel}
\Delta_q=\Delta_{1-q},
\end{equation}
 with $\Delta_q=\tau_q-q+1$. For 
the singularity spectrum it gives $f(2-\alpha)=f(\alpha)+1-\alpha$.

The validity of these different relations will be investigated on two particularly simple models of quantum maps that we describe in the next section.

\section{Models}

\subsection{Intermediate map}
The properties of section \ref{knownfacts} have been first observed for
wave functions in the 3D Anderson model at criticality. 
In the present paper the first model we will consider is a quantum map whose eigenfunctions
display similar multifractal properties in momentum representation. It
corresponds to a quantization of a classical map, defined on
the torus by
\begin{equation}
\bar{p}=p+\gamma \;\mbox{(mod}\;\mbox{1)}\;;\;\;\;
\bar{q} = q+ 2\bar{p} \;\mbox{(mod}\;\mbox{1)}\;,
\end{equation}
where $p$ is the momentum variable and $q$ the angle variable, while
$\bar{p}$ and $\bar{q}$ are the same quantities after one iteration of 
the map.  
The corresponding quantum evolution can be expressed as 
\begin{equation}
\bar{\psi} = \hat{U} \psi= e^{-2i\pi\hat{p}^2/N} e^{2i\pi \gamma \hat{q}}\psi
\end{equation}
in operator notation, or equivalently as 
a $N \times N$ matrix in momentum representation:
\begin{equation}
\label{intermediate_map}
U_{kk'}=\frac{\exp(-2i\pi k^2/N)}{N}
\frac{1-\exp(2i\pi N \gamma)}{1-\exp (2i\pi (k-k'+N\gamma)/N)},
 \end{equation}
with $0\leq k,k'\leq N-1$ \cite{giraud}. 
For generic irrational $\gamma$, the spectral statistics of $\hat{U}$
are expected to follow Random Matrix Theory.
When $\gamma$ is a rational number, $\gamma=a/b$ with $a,b$ integers,
spectral statistics are of
intermediate type and eigenvectors of the map display
multifractal properties.
 In order to study the effect of ensemble averaging on
multifractal exponents, we will consider a random version introduced in 
\cite{bogomolny}, where the phases $2\pi k^2/N$ are replaced by independent
random phases $\phi_k$. We will also present numerical results obtained for
the initial map given by \eqref{intermediate_map}.

\subsection{Anderson map}

An important system where multifractal wave functions have been observed
corresponds to electrons in a disordered potential in dimension three.
Indeed, the Anderson model \cite{mirlinRMP08} which describes such
a situation displays a transition between a localized phase 
(exponentially localized wave functions) and
a diffusive phase (ergodic wave functions) 
for a critical strength of disorder.  At
the transition point, the wave functions display multifractal properties 
\cite{mirlinRMP08} and
the spectral statistics are of the intermediate type \cite{braun}.
In order to compare this type of system with the previous one, 
we have studied a one-dimensional system with incommensurate
frequencies, which has been shown to display an Anderson-like transition
\cite{dima}.  In \cite{delande} it was shown that it can be implemented with cold atoms in an optical
lattice, which enables to probe experimentally the
Anderson transition.
The system is a generalization of the quantum
kicked rotator model, and is described by a unitary operator 
which evolves the system over one time interval:

\begin{equation}
\label{qmap}
\bar{\psi} = \hat{U} \psi =  e^{-iV(\hat{\theta},t)}
 e^{-iH_0(\hat{n})} \psi,
\end{equation}
with $V(\hat{\theta},t)= k(1+\epsilon\cos{\omega_1 t}\cos{\omega_2 t})
\cos{\hat{\theta}}$ (here time $t$ corresponds to number of kicks). Here $\omega_1$ and $\omega_2$ should be two frequencies mutually incommensurate.
Following
\cite{delande} we chose in the
simulations $\epsilon=0.75$, $\omega_1=2\pi\lambda^{-1}$ and $\omega_2=2 \pi\lambda^{-2}$, where $\lambda=1.3247\ldots$ is the real root of the cubic equation $x^3-x-1=0$. In \cite{dima} it was shown that this system displays an
Anderson-like transition at the critical value $k_c\approx 1.81$, but multifractality of this system was up to now not verified.
 The function $H_0(n)$ can be chosen either
by taking $H_0(n)=n^2/2$ (free evolution) or as in the preceding case
by replacing this quantity by independent random phases uniformly distributed in the interval $[0,2\pi[$.
This is actually what we chose to do in our numerical simulations,
in order to increase the stability and accuracy of the numerical results.

As a wave packet spreads slowly at the transition, one has to iterate
the map for a long time in order to obtain data on a sufficiently large
wave function.  We found that in order to reach vector sizes of order 
$2^{11}$, it was necessary to iterate the map up to $t=10^{8}$.  Such values
are not realistic for experiments with cold atoms (limited to a few hundreds
kicks) but allow to obtain more precise results. 

\subsection{Variations on the models}

In both models the evolution of the system during one time step
has the form of an operator diagonal in momentum 
(kinetic term) times an operator
diagonal in position (kick term).  In both cases,
it has been common practice in the field to replace  the
kinetic term by random phases. This allows to obtain a similar
dynamics but with a more generic behavior.  Moreover, it enables
averages over random phases to be performed, which makes
numerical and analytical studies more precise.  In contrast, in experiments
with cold atoms it is easier to perform iterations with a true
kinetic term rather than with random phases. In order to assess the 
effect of this modification, we will therefore
use both approaches in the study of multifractal
properties of wave functions.


Many works on multifractal wave functions have focused on eigenstates
of the Hamiltonian (see e.g. \cite{mirlinRMP08}).  For the intermediate map,
the evolution operator has eigenvectors which can be numerically found
and explored. It is also possible to evolve wave packets, for example
initially concentrated on one momentum state, and to study the multifractality
of the wave packet as time increases.  This corresponds
more closely to what can be explored in experiments.
For the intermediate map, this
process can be understood as the dynamics of
a superposition of eigenvectors of the evolution operator.
However, in the case of the Anderson map, the evolution operator is 
time dependent (the continuous time problem being not periodic),
and the connection with eigenvectors is lacking.  In the following,
we will explore and compare the multifractality
of both eigenfunctions and time-evolved wave packets for the intermediate map.

\section{Numerical methods}

As is well-known, the numerical estimation of multifractal dimensions is
very sensitive to finite-size effects. In the present work we have
analyzed and compared different numerical methods in order to compute
accurately the multifractal exponents. In this section we briefly review the
methods we used.

\subsection{Box-counting method}

The most straightforward method is to compute directly
the moments of the wave functions through the scaling of the moments
\eqref{moments} given by $P_q= \sum_i |\psi_i|^{2q}$. If the scaling
\eqref{scaling} holds true, then $\log\langle P_q\rangle$ is a linear
function of $\log N$ and its slopes yield the exponents $\tau_q$.
For $q<0$, coarse-graining over neighboring sites is necessary in order to 
avoid instabilities due to very small values of $|\psi_i|$.


A variation of the moment method is the box-counting method. It
consists in taking a vector of fixed size
$N$ and summing the  $|\psi_i|^2$ over boxes of increasing length. 
If $N=2^n$, then we define
\begin{equation}
\pqk= \sum_{i=0}^{2^{n-k}-1}\left(\sum_{j=0}^{2^k-1}|\psi_{i 2^k+j}|^2\right)^q
\end{equation}
which corresponds to averaging the measure over boxes of length $2^k$,
$0\leq k\leq n$. Starting from $k=2$ allows to smooth out exceedingly tiny
values of $|\psi_i|^2$ which otherwise yield non-accurate estimates of $D_q$ for
negative $q$. The two methods above give similar results, as was also carefully checked for the Anderson model in
\cite{deltaq}.  We will therefore
in the following present results using the box-counting method to assess the
properties of this type of procedures.

\subsection{Wavelet transform method}
Recently, alternative procedures to compute the multifractal spectrum based on the
wavelet transform   
were developed \cite{arneodo}. The wavelets form a basis of functions as does the Fourier
basis and a function expressed in this new basis gives the wavelet transform
(WT). Unlike the Fourier basis wavelets are localized both in position and 
in momentum space (or time and frequency space). They are therefore suitable
to probe the local variations of a function at different scales. They have
become an essential tool for image and sound processing and compression. 

A wavelet basis is constructed from a single function $g$ called analyzing or mother wavelet. The rest of 
the basis is constructed by translations and compressions (expansions) of
the analyzing wavelet $g$. The translations define a space variable while
the compressions define the scale at which the function is analyzed. We define the
WT of a (real) function $h$ as
\begin{equation} 
\label{wave}
T_h(A,B)=\frac{1}{A}\int dx\, h(x)\;g\left(\frac{x-B}{A}\right),
\end{equation}
where $A$ is the scale variable and $B$ is the space variable.
As a consequence, $T_h(A,B)$ can be interpreted as  a measure of how close
the function $h$ is to the mother wavelet at point $B$ and at scale $A$.

The $\tau_q$ can be extracted from the WT in the following
 way. We define the partition function 
\begin{equation}
{\cal Z}(q,A)=\sum_{B_i}|T_h(A,B_i)|^q,
\end{equation}
where $(A,B_i)_i$ are the local maxima at scale size $A$ and $q$ is
real. It can be shown that $\tau_q$ appears as the exponent in the power law
behavior of ${\cal Z}(q,A)$ 

\begin{equation}\label{powlaw}
{\cal Z}(q,A)_{
\begin{smallmatrix}
\mbox{\large $\sim$}\ \\ \mbox{\tiny $A\to 0^+$}
\end{smallmatrix}
}A^{\tau_q}.
\end{equation}
This is essentially the method known as wavelet transform modulus maxima (WTMM) \cite{arneodo}. This method is developed for continuous wavelet function.  

If the function $h$ is sampled as an $N$-dimensional vector with $N=2^n$ the wavelet transform can be discretized and implemented efficiently  by a hierarchical algorithm \cite{mallat} resulting in a fast wavelet transform, (FWT). The scale and space parameters take the values
\begin{eqnarray}
A&=&1,\frac{1}{2},\frac{1}{4},\ldots,\frac{1}{2^{n-1}}\\
B_A&\in&\left\{1,2,\ldots,\frac{1}{A}\right\}
\end{eqnarray}
respectively. 
Starting from a wave function $\psi$ and
using a proper normalization at each scale, we redefine the partition function as
\begin{equation}
\label{ZarneoMM}
{\cal Z}(q,A)=\sum_{B_i}\left[\frac{|T_{|\psi|^2}(A,B_i)|}
{\sum _{B_i}|T_{|\psi|^2}(A,B_i)|}\right]^q
\end{equation}
where again $(A,B_i)_i$ are the local maxima at scale $A$. As in the continuous case ${\cal Z}(q,A)$ exhibits the same power law behavior as Eq.~(\ref{powlaw}). In the following, we will present results using this implementation of the wavelet method, using the Daubechies 4 mother wavelet \cite{daub}.

We note that, although the partition function (\ref{ZarneoMM}) is the most
standard, recently it was numerically observed \cite{ignacio} that for a complex multifractal wave function $\psi$, the exponents $\tau_q$ can also be obtained from the power law behavior of a modified partition function defined from the complex FWT of $\psi$ and using $2q$ as exponents in Eq.~(\ref{ZarneoMM}).

\section{Results}

\subsection{Numerical computation of the multifractal exponents}

\begin{figure}[h]
\includegraphics[width=\linewidth]{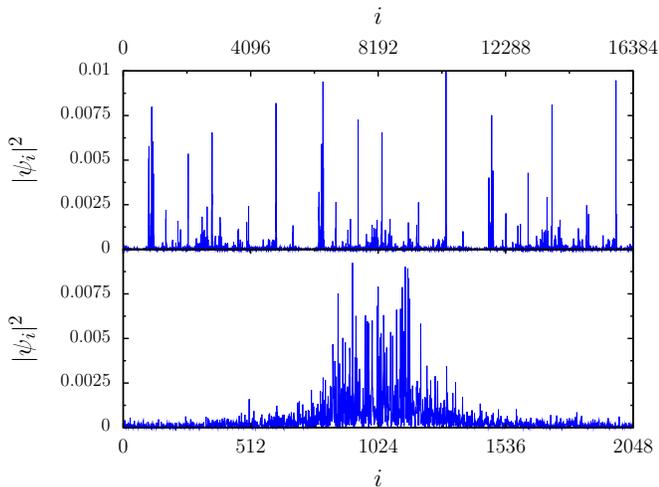}
\caption{(Color online) Top panel: Instance of an eigenvector of the intermediate map with random phases for $N=2^{14}$ and $\gamma=1/3$. Bottom panel: Instance of an iterate of the Anderson map with random phases for $k=1.81$, $N=2^{11}$ and $t=10^8$; $|\psi(0)\rangle=|i\rangle$ with $i=1024$.
\label{fig:psi2}}
\end{figure}

\begin{figure}[h]
\includegraphics[width=\linewidth]{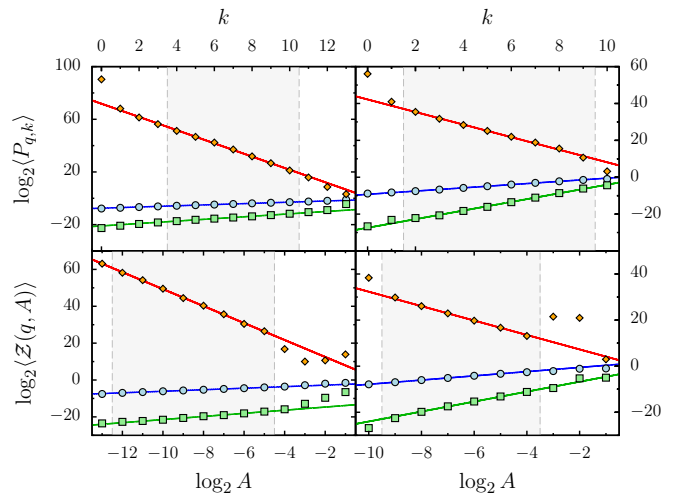}
\caption{(Color online) Top panels: Average moments $\log_2 \langle P_{q,k}\rangle$ as a function of $k$, the logarithm of the box sizes.
Bottom panels: Partition function $\log_2 \langle {\cal Z}(q,A)\rangle$ as a function of the logarithm of the scale $A$. The left panels correspond to eigenvectors of the Intermediate map with random phases, $N=2^{14}$, and $\gamma=1/3$. Here the average is taken over $98304$ vectors (resp.\ $32768$ vectors for the partition function). The right panels correspond to iterates of the Anderson map for $k=1.81$, $N=2^{11}$, $|\psi(0)\rangle=|i\rangle$ with $i=1024$ and $t=10^8$, where the average is taken over $1302$ vectors. The values chosen for $q$ are : $q=-2$ (red diamonds), $q=2$ (blue circles), and $q=6$ (green squares). The gray shaded regions show the fitting interval.
\label{fig:LogZq}}
\end{figure}

In this subsection, we present numerical results corresponding to the
multifractal exponents for the intermediate quantum
map model (\ref{intermediate_map}) and the Anderson map (\ref{qmap}).  Examples of wave functions for both models are
shown in Fig.~\ref{fig:psi2}. The two sets of multifractal dimensions
$D_q$ and $\dqtyp$ were computed from $\log_2 \langle \pqk\rangle$ and $\langle \log_2 \pqk \rangle$
respectively for the box-counting method, and from $\log_2 \langle\calz(q,A)\rangle$ and 
$\langle \log_2\calz(q,A)\rangle$ respectively for the wavelet method. For the
Anderson map, as less realizations of the random phases could be computed, the same set of random realizations
were used for the two methods investigated in order to ensure that comparable quantities were plotted.
Figure~\ref{fig:LogZq} illustrates the scaling of these quantities for the
two models considered. It displays $\log_2 \langle \pqk\rangle$ as a function
of the logarithm of the box size (top) and 
$ \log_2\langle\calz(q,A)\rangle$ as a function of the logarithm of the
scale parameter $\log_2 A$ (bottom) for different values of $q$. 
We chose to show the scaling of these moments since it is the worst case,
the curves for $\langle \log_2 \pqk \rangle$ and $\langle \log_2\calz(q,A)\rangle$ being always closer to linear functions. Nevertheless, Fig.~\ref{fig:LogZq} shows that the scaling is indeed
 linear over several orders of magnitude. The slopes of the linear fits give  
the multifractal exponents. In the two methods, there is a certain freedom
in determining the range of box sizes (or scales for the wavelet method)
over which the linear fit is made (the one we chose is indicated by the shaded area in Fig.~\ref{fig:LogZq}).
Usually for moderate values of $q$ in absolute value, the result does not
depend very much on this choice.
As can be seen in Fig.~\ref{fig:LogZq}, the data are well fitted by
a linear function in the range chosen; however, there is still an uncertainty 
on the slope, which usually gets larger for large negative $q$.  In the next figures of this subsection,
the uncertainty of the linear fit for the set of points chosen is plotted together with the mean value,
in order to give an estimate of the reliability of the values obtained.

\begin{figure}[h]
\begin{center}
\includegraphics[width=\linewidth]{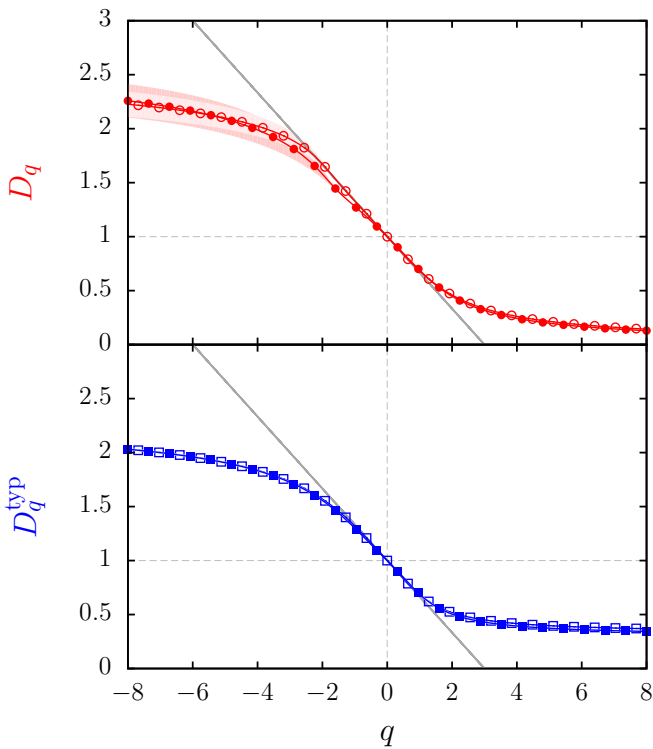}
\end{center}
\caption{(Color online) Top panel: Multifractal exponents $D_q$ for eigenvectors of the Intermediate map with random phases, $N=2^{14}$, and $\gamma=1/3$. Empty/filled red circles correspond to the method of moments/wavelets. The (light blue, light red) shaded regions indicate standard error in the least squares fitting. The multifractal analysis was done over $98304$ vectors with box sizes ranging from $16$ to $1024$ and scales ranging from $2-12$ to $2-5$. Bottom panel: Typical multifractal exponents $D^{\rm typ}_q$ for the same data. In both panels, the gray solid line is the linear approximation $1-q/3$.\label{fig:Dqb}}
\end{figure}

\begin{figure}[h]
\begin{center}
\includegraphics[width=\linewidth]{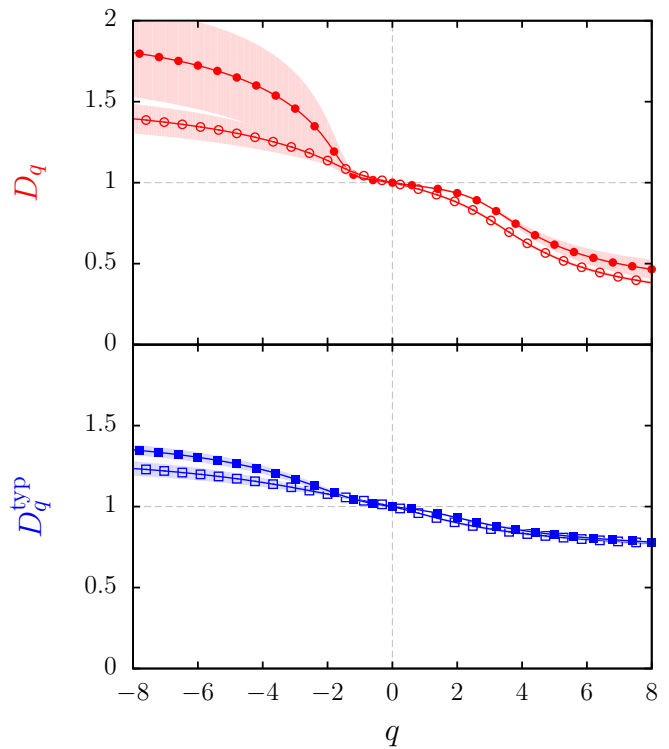}
\end{center}
\caption{(Color online) Multifractal exponents $D_q$ and $D^{\rm typ}_q$ for iterates of the Anderson map with $k=1.81$, $N=2^{11}$. 
Same convention as in Fig.~\ref{fig:Dqb}. The multifractal analysis was done over $1302$ realizations of size $N=2^{11}$ with box sizes ranging from $4$ to $512$ and scales ranging from $2-9$ to $2-4$. \label{fig:DqAnd}}
\end{figure}

The values of $D_q$ and $\dqtyp$ as a function of $q$ are
 presented in Fig.~\ref{fig:Dqb} for the random intermediate map and in 
Fig.~\ref{fig:DqAnd} for the Anderson map. In both cases, we observe a spectrum
typical of multifractal wave functions.
In the intermediate case,  we observe that the two methods give comparable
results with small uncertainty, although it gets larger for large
negative $q$.   For the Anderson map, the uncertainty gets very large
for $q \lesssim -2$, and besides it begins to depend strongly on the range of
box sizes (box-counting method) or scales (wavelet method) chosen (data not shown).
We attribute the larger uncertainty for the wavelet method to the seemingly stronger sensitivity of
 this method to exceptionally small values. 
We note that for the Anderson map only a vector size of $2^{11}$ could be reached numerically
because of the computational power required to iterate the Anderson map for long times.
The discrepancy between
 the two methods is smaller for $\dqtyp$, which can be
 understood by the fact that  $\langle\pqk \rangle$ or
 $\langle\calz(q,A)\rangle$ are more sensitive to rare events than $\langle
  \log_2 \pqk \rangle$ and $\langle \log_2\calz(q,A)\rangle$.  

Our data nevertheless show that the iterates of the Anderson map
display multifractal behavior.  In view of the recent implementation 
of such maps with cold atoms \cite{delande}, this indicates that
in principle one can observe multifractality of this map in cold atom experiments; we note that in \cite{delandebis} properties of the wave functions of
this experimental system were investigated, but focused on the envelope of
the wave packet.

In both the intermediate map and the Anderson map, $\dqtyp$
goes to a constant $\alpha_{+}$ for large $q$, 
which corresponds to the fact already
mentioned that $\tauqtyp$ is expected to behave as $\tauqtyp=q\alpha_{+}$
for $q>q_{+}$. This will be studied in more detail in the next section.

In~\cite{MGG} we observed the existence of a linear regime 
around $q=0$, with slope $-1/b$ for the random intermediate map with
 parameter $\gamma=1/b$. This regime seemed to be valid in quite a large
 interval around zero. 
The data displayed in Fig.~\ref{fig:Dqb} obtained by two
 different large-scale computations confirm this result, and show that this
 regime exists for both types of averages. We will come back to this
 property in section \ref{linear_regime}.

\begin{figure}[h]
\begin{center}
\includegraphics[width=\linewidth]{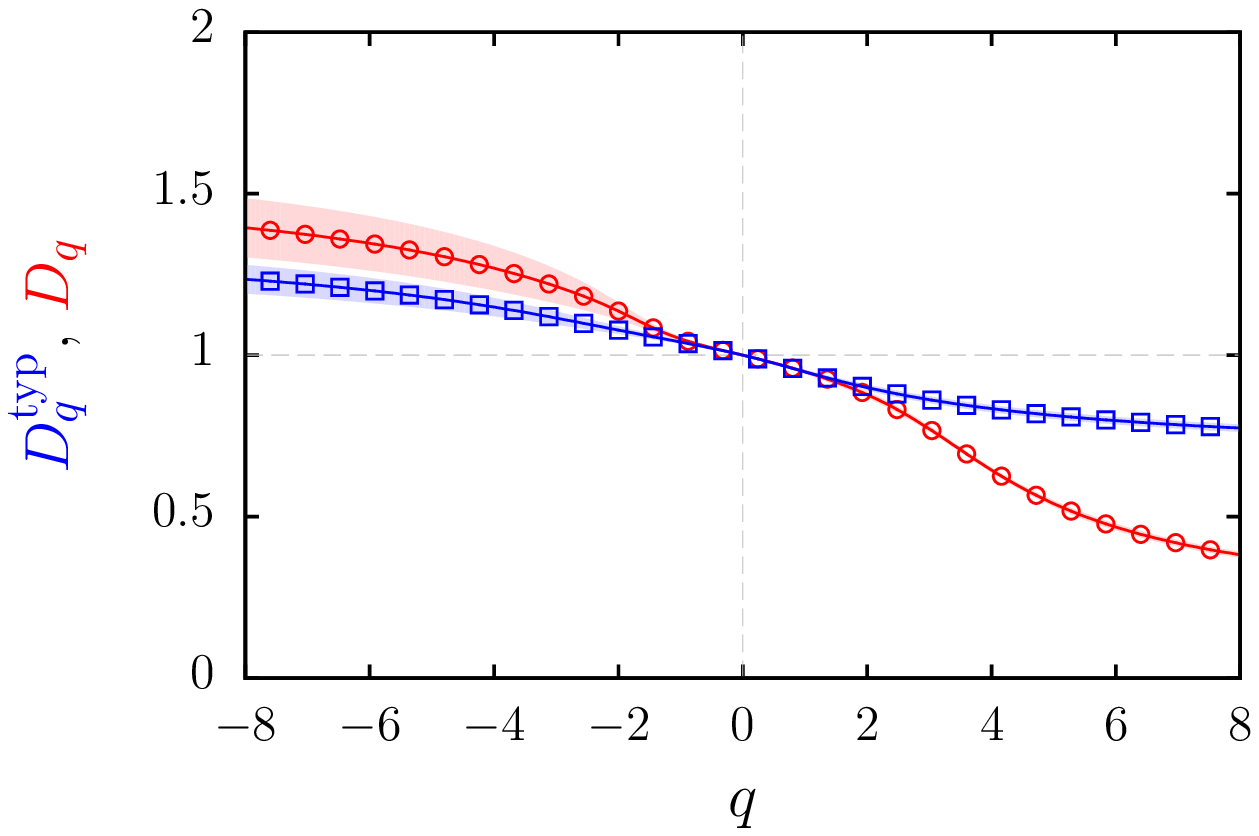}\\\vskip 10pt
\includegraphics[width=\linewidth]{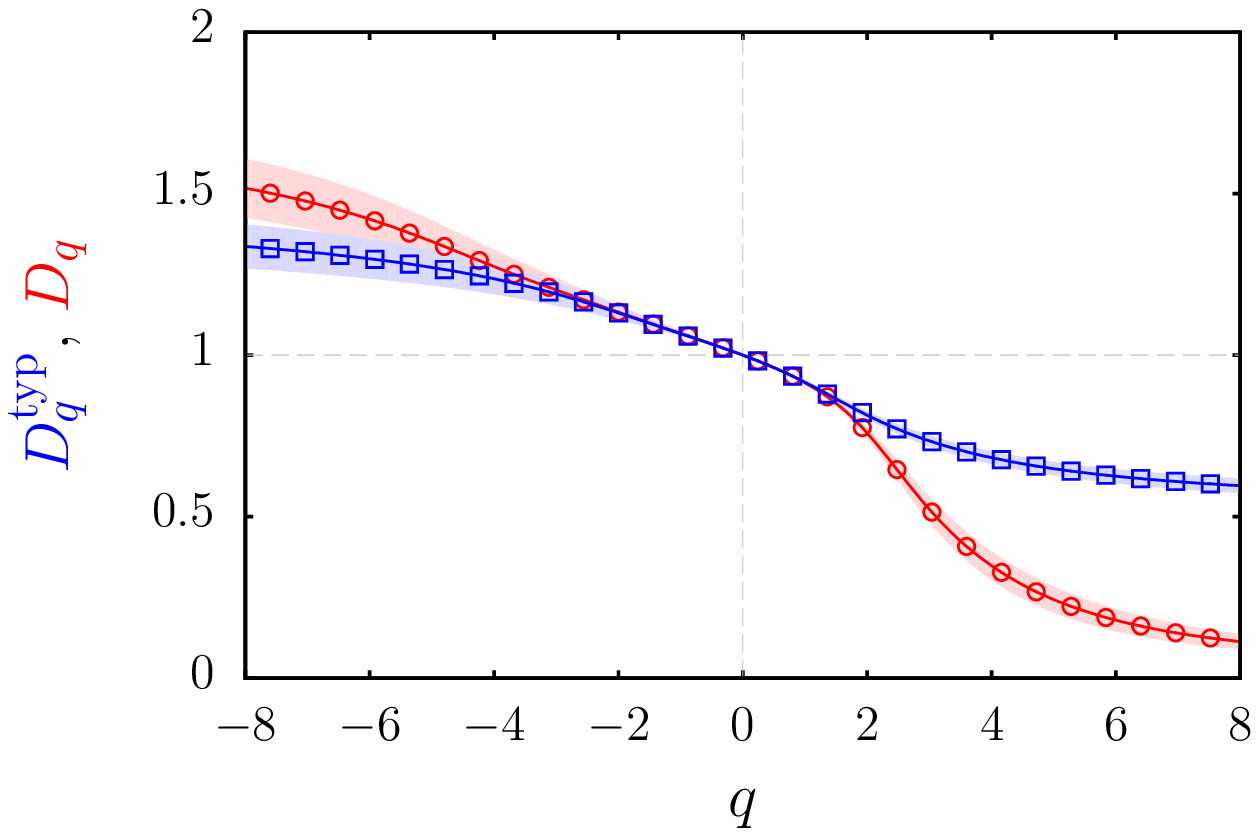}
\end{center}
\caption{(Color online) Top: Multifractal exponents $D_q$ and $D^{\rm typ}_q$ for the Anderson map with $k=1.81$, $N=2^{11}$, $|\psi(0)\rangle=|i\rangle$ with $i=1024$ and $t=10^8$. Same convention as in Fig.~\ref{fig:Dqb}. The multifractal analysis was done over 1302 realizations with box sizes ranging from 4 to 512. Bottom:
Multifractal exponents $D_q$ and $D^{\rm typ}_q$ for iterates of the intermediate map with $b=3$, $N=2^{12}$, $|\psi(0)\rangle=|i\rangle$ with $i=2048$ and $t=10^8$. The multifractal analysis was done over 2000 realizations with box sizes ranging from 4 to 512. In both cases we used the box-counting method.
\label{fig:DqDqtypit}}
\end{figure}

The data presented in Fig.~\ref{fig:DqAnd} show that the Anderson vectors are
less fractal than the eigenvectors of the intermediate map for $b=3$.
This can be expected from the fact that Anderson vectors correspond to iterates of a
 quantum map, which in general are less fractal than eigenvectors \cite{girgeor}.  In order to compare similar quantities, we display in
 Fig.~\ref{fig:DqDqtypit} multifractal dimensions for vectors obtained by
 iteration of the intermediate map for $b=3$. One sees that using iterates instead
of eigenvectors also in this case reduces the overall multifractality at a given $q$. Although a simple relation between multifractal exponents of eigenvectors and
iterates is still lacking, this indicates that it might be possible to use 
experimental results from cold atom experiments to infer properties of
the eigenvectors of the 3D Anderson transition.

\begin{figure}[h]
\begin{center}
\includegraphics[width=\linewidth]{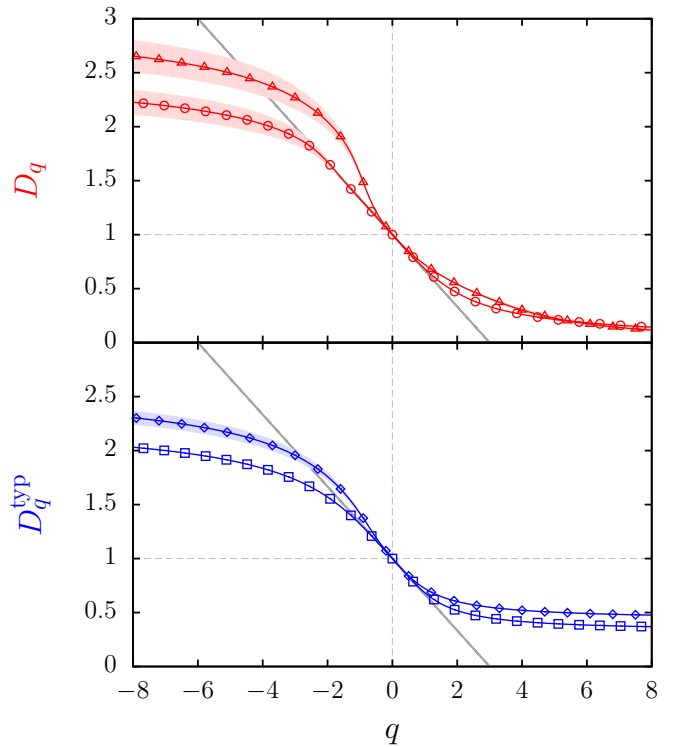}
\end{center}
\caption{(Color online) Top panel: Multifractal exponents $D_q$ for eigenvectors of the intermediate map with (circles)
and without (triangles) random phases. 
 Bottom panel: Typical multifractal exponents $D^{\rm typ}_q$ for eigenvectors of the intermediate map with (squares) and without (diamonds) random phases. 
In both cases we used the box-counting method. Conventions and parameter values are the same as in Fig.~\ref{fig:Dqb} except that for non-random phases, a single realization of size $N=2^{14}$ was considered. In both panels, the gray solid line is the linear approximation $1-q/3$.\label{fig:Dqbnonrand}}
\end{figure}

In order to assess the effect of random phases,
 we present in Fig.~\ref{fig:Dqbnonrand} the results of numerical computation of
$D_q$ and $\dqtyp$ for the intermediate map \eqref{intermediate_map} without
random phases. Although spectral statistics for random and non-random vectors are
very close \cite{bogomolny}, the obtained fractal dimensions are quite
different. Moreover it can be seen that there still exists a difference between
$D_q$ and $\dqtyp$, although the map is not random any more. 
The discrepancy in the two sets of exponents was to our knowledge 
up to date observed
only in disordered systems, where rare events are created by specific
realizations of disorder.  It is interesting to see that this effect can
be observed in a dynamical system without any disorder whatsoever. 
This
discrepancy is due to the fact that the average in
Eqs.~\eqref{scaling}-\eqref{scalingtyp} is performed over several
eigenvectors of a single realization of the map, which gives a certain
dispersion of the moment distribution. The average over eigenvectors in the
intermediate map thus suffices to create the effect even in a deterministic
map.  Therefore one can also observe the separation between the two sets
of exponents in quantum systems
without disorder.  The rare events in this case correspond to rare eigenvectors
of the evolution operator having large moments.

The numerical results displayed in this Subsection indicate that multifractality is indeed present in all
models considered.
Furthermore, our results show that
for an appropriate choice of a range of box sizes (box-counting method) or scales 
(wavelet method), both methods are based on curves well-fitted
by linear function over a wide interval, with a small uncertainty on the exponent extracted.
We believe that the good agreement for $q \gtrsim -2$ between the two methods, and the small
uncertainty found for the linear fit, indicates that our numerically extracted multifractal exponents
are reliable (up to finite size effects).  For $q \lesssim -2$, the numerical uncertainty increases
for decreasing $q$, and the two methods give increasingly different results.
The results presented here show that our data are still reliable, although less precise, for eigenvectors of the intermediate map, even for negative $q$.
However, in the case of the Anderson map, the uncertainty is too large to give reliable
results for $q \lesssim -2$.  This is certainly due to the fact that both the number
of realizations and the vector sizes are smaller in this case, which makes it difficult to find reliable results in the more demanding regime of large
negative $q$.    In the regime of $q \gtrsim -2$, our results indicate that
the two methods can be used and give similar results.  In the
more demanding cases (large negative $q$) we found the box-counting method more reliable and accurate, and therefore the numerical results presented in the following
Subsections correspond to this method.

\subsection{Linear regime}
\label{linear_regime}

In \cite{MGG} the first investigations of the multifractal exponents for
the random intermediate map showed the presence of a linear regime 
around $q=0$ over a relatively large range of $q$ values.
We recall that the map \eqref{intermediate_map} displays multifractal
properties of rational values of the parameter $\gamma=a/b$.
Based on semi-heuristic arguments, this linear regime was shown
to be described by:
\begin{equation}
\label{DQlin} D_q \approx D_q^{\textrm{lin}}=1-\frac{q}{b}.
\end{equation}
The relation (\ref{DQlin}) enables to link the spectral statistics
and the distribution of $D_q$ around $q=0$ in a systematic way since
both are controlled by the parameter $b$ explicitly.  While
one expects that some form of 
linear regime should exist over small
intervals for any smooth curve, the extent of it in this particular
model indicated a small second derivative near $q=0$.  This feature
was seen in the PRBM model \cite{mirlinRMP08} but in the regime of
weak multifractality where the derivative at $q=0$ of $D_q$ is
very close to zero.  

\begin{figure}[h]
\begin{center}
\includegraphics[width=\linewidth]{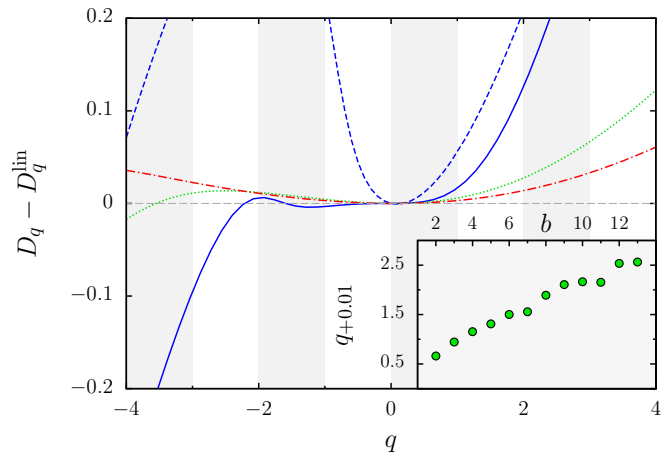}
\end{center}
\caption{(Color online) Difference $D_q-D_q^{\textrm{lin}}$ for eigenvectors of the
  intermediate map with random phases for $\gamma=1/3$ (blue solid curve), $\gamma=1/7$ (green dotted curve) and $\gamma=1/11$ (red
  dash-dotted curve). Blue dashed
  curve shows the same difference for the intermediate map without random
phases for $\gamma=1/3$. Inset: separation points between $D_q$ and
$D_q^{\textrm{lin}}$ (green dots), determined by  $(D_q-D_q^{\textrm{lin}})/(D_q+D_q^{\textrm{lin}})=0.01$, for eigenvectors of the intermediate map with random phases for different values of $\gamma=1/b$.
Other parameters as in Fig.~\ref{fig:Dqbnonrand}. \label{figE1} }
\end{figure}

Figure \ref{figE1} displays the extent of the linear
regime for the intermediate map for three values of the parameter 
$\gamma=1/b$.
The data presented show that the linear regime is present in all three cases,
although its extent seems to be larger for large $b$ (weak multifractality).
 This indicates
that the linear regime is a robust feature of the random intermediate
map.  To explore more precisely the dependence of this regime on the value
of $b$, the inset of Fig.~\ref{figE1} shows the separation point
between the actual exponent and the linear value, defined by
a constant relative difference (set to $1\%$), for all values of $b$
between $b=2$ and $b=13$. The data presented
show that indeed the extent of the linear regime grows with $b$, although
the precise law of this growth is difficult to specify.

These results correspond to the random intermediate map, where the
kinetic term is replaced by random phase.  It is important to explore also
the behavior of the deterministic intermediate map, where
the kinetic term is kept as a function of momentum.
In this case,
Fig.~\ref{figE1} shows the presence of a much smaller linear regime. 
This indicates a strong difference between the
random model and the deterministic one, and that a large linear regime
is a property restricted to a certain kind of models.

\begin{figure}[h]
\begin{center}
\includegraphics[width=\linewidth]{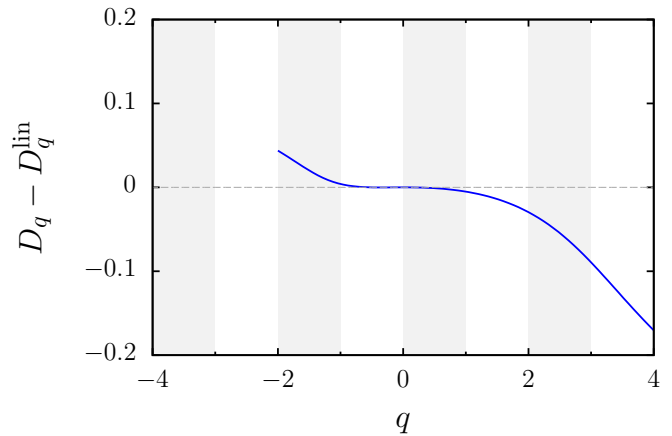}
\end{center}
\caption{(Color online) Difference $D_q-D_q^{\textrm{lin}}$ for iterates of
  the Anderson map, with $D_q^{\textrm{lin}}$ defined by
$D_q^{\textrm{lin}}=1+qD_q'(0)$.
 Same parameters as in Fig.~\ref{fig:DqAnd}. \label{figE3} }
\end{figure}

The data for the Anderson map with random phases, shown in Fig.~\ref{figE3},
also show a linear regime, comparable with the random intermediate map.
Again, as the data correspond to iterates of wave packets, not 
eigenvectors, the multifractality is weaker than for other simulations
of the Anderson model using eigenvectors \cite{mirlinRMP08}. This might explain
why the linear regime that is visible seems larger than for Anderson transition eigenstates.

\subsection{Average vs typical multifractal exponents}

\begin{figure}[h]
\begin{center}
\includegraphics[width=\linewidth]{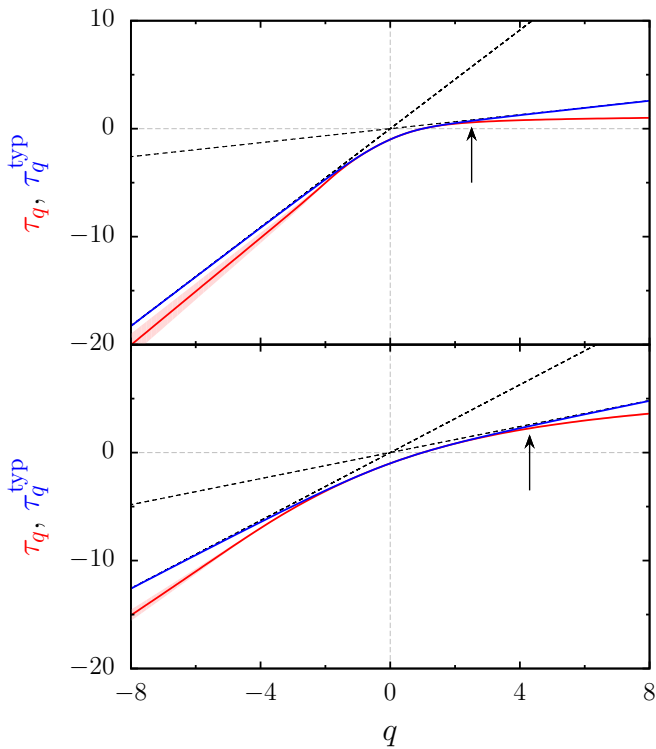}
\end{center}
\caption{(Color online) Top panel: Exponents $\tau_q$ (red solid lower curve)
  and $\tau_q^{\rm typ}$ (blue solid upper curve) as a function of $q$ for
  eigenvectors of the intermediate map with random phases, $N=2^{14}$, and
  $\gamma=1/3$. Bottom panel: same but for $\gamma=1/11$. In both panels, the
  black dashed lines are linear fits of $\tau_q^{\rm typ}$ at large $|q|$. The
  arrows indicate the critical $q$ value determined by $x_{q}=1$ (see
  Subsection V.D), equal to $2.51$ for $\gamma=1/3$ and to $4.30$ for $\gamma=1/11$. The slopes of the black dashed lines are respectively given by $\alpha_-\approx 2.28$ and $\alpha_+\approx 0.32$ for $\gamma=1/3$ and by $\alpha_-\approx 1.57$ and $\alpha_+\approx 0.60$ for $\gamma=1/11$. \label{fig:tauq}} 
\end{figure}

In disordered systems, the statistical distribution of the moments of
the wave function is responsible for a discrepancy between the
multifractal exponents $D_q$ and $\dqtyp$ calculated respectively by
averaging over the moments themselves or over their logarithm. 
The two sets of exponents are expected to match only in some
region $q\in[q_-,q_+]$. Outside this range, $\tauqtyp$ should follow a linear
behavior. Figure \ref{fig:tauq} displays results for $\tauq$ and $\tauqtyp$ for
the intermediate map with parameter $\gamma=1/b$ for two representative values of $b$. As
expected from the PRBM model \cite{mirlinRMP08}, in the case of weak multifractality ($\gamma=1/11$, bottom panel of
Fig.~\ref{fig:tauq}), the range over which the exponents are equal is
wider than for strong multifractality ($\gamma=1/3$, top panel of
Fig.~\ref{fig:tauq}) . Beyond that interval
the behavior of $\tauqtyp$ is linear for both values of $\gamma$. For positive $q$ the linear tail appears around $q\simeq
2.5$ for $\gamma=1/3$ and $q\simeq 4.3$ for $\gamma=1/11$. The slopes of the linear tails give $\alpha_-$ and $\alpha_+$.
According to the theory (see Section II), these values of $\alpha$
  correspond to the terminating point of the singularity
  spectrum $\ftyp(\alpha)$ defined in Eq.~\eqref{falpha}. We obtained comparable results
for the Anderson map (data not shown).

\begin{figure}[h]
\begin{center}
\includegraphics[width=\linewidth]{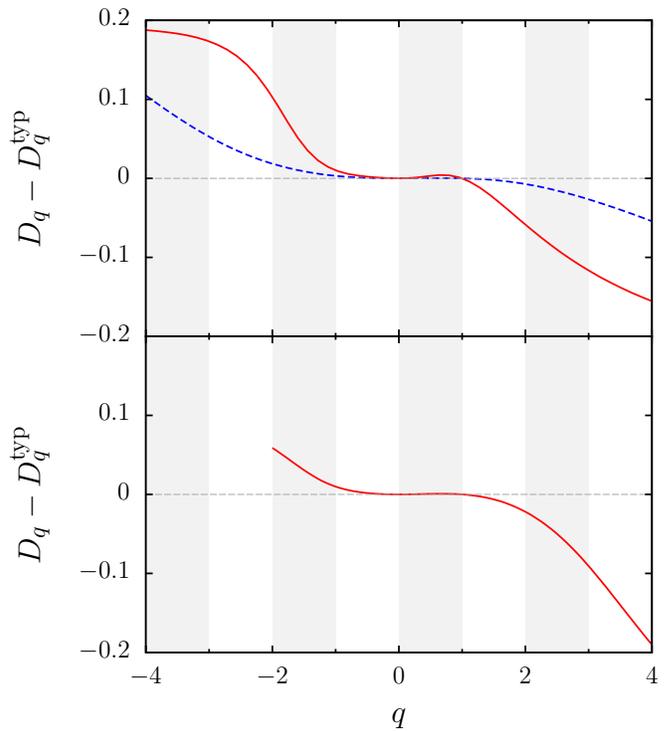}
\end{center}
\caption{(Color online) Top panel: Difference $D_q-D_q^{\rm typ}$ between average and typical exponents for eigenvectors of the intermediate map with random phases for $\gamma=1/3$ (red solid curve) and  $\gamma=1/11$ (blue dashed curve). Bottom panel: same figure for iterates of the Anderson map. Same parameters as in Fig.~ \ref{fig:DqAnd}. \label{fig:DeltaDq} }
\end{figure}

\begin{figure}[h]
\begin{center}
\includegraphics[width=\linewidth]{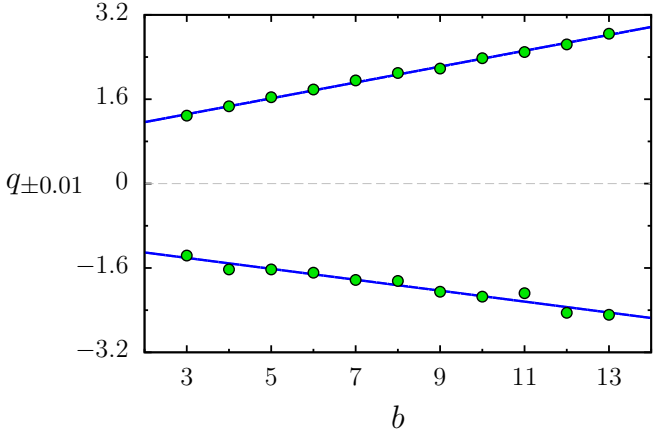}
\end{center}
\caption{(Color online) Separation points $q_{\pm 0.01}$ between $D_q$ and $D_q^{\rm typ}$ (green dots),
determined by  $(D_q-D_q^{\rm typ})/(D_q+D_q^{\rm typ})=\pm 0.01$, for eigenvectors of the intermediate map with random phases for different values of $\gamma=1/b$. The blue solid lines are the linear fits $q_{- 0.01}\approx -0.10 b-1.10$ and $q_{+ 0.01}\approx 0.15 b+0.86$. Same parameters as in Fig.~\ref{fig:DqAnd}. \label{figdist} }
\end{figure}

In order to take a closer look at the discrepancy between the two sets
of exponents, we plot the difference between the exponents
$D_q$ and $\dqtyp$ for both systems. It is clearly seen in Fig.~\ref{fig:DeltaDq} that
in all cases the regime where the exponents coincide is only about $-1<q<1$. At
this scale the separation between $\tau_q$ and $\tauqtyp$ occurs
around $q=1$. In order to obtain more systematically the separation point, we
have plotted in Fig.~\ref{figdist}  the value of $q$ defined by a constant relative value of the difference of the two exponents (set equal to $0.01$) for intermediate maps with different parameters 
$\gamma=1/b$; this allows
to get comparable data independently of the value of the exponents.
The results show a clear linear scaling of the separation point with respect
to $b$, obeying the formulas $q_{- 0.01}+1\approx-0.1(b+1)$ and
$q_{+ 0.01}-1\approx 0.15(b-1)$.  Changing the threshold of relative value
from $0.01$ to $0.02$ gives also a linear scaling, with different slope (data not shown).  

\begin{figure}[h]
\begin{center}
\includegraphics[width=\linewidth]{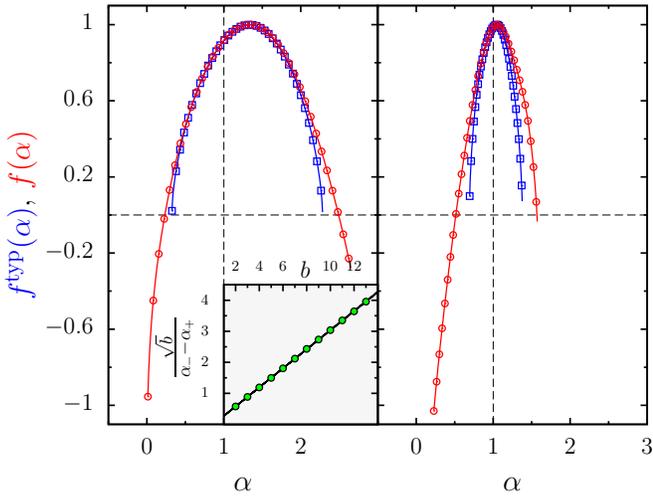}
\end{center}
\caption{(Color online) Left panel: Singularity spectra $f(\alpha)$ and
  $\ftyp(\alpha)$ for eigenvectors of the intermediate map with random
  phases for $\gamma=1/3$. The inset shows the linear behavior of
  $\sqrt{b}/(\alpha_--\alpha_+)\simeq 0.307 b-0.037$ as a function of
  $b$ (in $\gamma=1/b$). The values of $\alpha_\pm$ have been
  extracted from linear fits of $\tau_q^{\rm typ}$ at large
  $|q|$. Right panel: Singularity spectra $f(\alpha)$ and
  $\ftyp(\alpha)$ for iterates of the Anderson map. In both panels,
  the singularity spectra are given in the range $\alpha(q=-16)\leq
  \alpha\leq \alpha(q=16)$. \label{fig:fa}} 
\end{figure}

The singularity spectrum $f(\alpha)$ defined in Eq.~\eqref{falpha}
is an alternative way to analyze multifractality and the discrepancy between
the two sets of multifractal exponents. In
Fig.~\ref{fig:fa} we show the singularity spectrum obtained for 
the intermediate
map \eqref{intermediate_map} with random phases (in \cite{MGG}
similar curves were obtained using directly the box-counting method to 
compute $f(\alpha)$; here we use the Legendre transform of the
exponents, obtaining similar results). As expected $\ftyp(\alpha)$
terminates at points $\alpha_{+}$ and $\alpha_{-}$ given by the
large-$q$ slopes of $\tauqtyp$, while  $f(\alpha)$ takes values below
zero coming from statistically rare events. In \cite{MGG}, it was shown
that a linear
approximation for $D_q$ yields a parabolic approximation for
$f(\alpha)$, giving in turn a behavior
$\alpha_{-}-\alpha_+\simeq1/\sqrt{b}$. We checked this behavior for all
values of $b$ between $b=2$ and $b=13$, thus confirming
the validity of this law, even beyond the linear regime (see inset of
Fig.~\ref{fig:fa}). Figure~\ref{fig:fa} allows a more direct
comparison between multifractality in the intermediate map and the
Anderson map: the narrower $f(\alpha)$ curve for Anderson corresponds
to a weaker multifractality.

\subsection{Moment distribution}

\begin{figure}[h]
\begin{center}
\includegraphics[width=\linewidth]{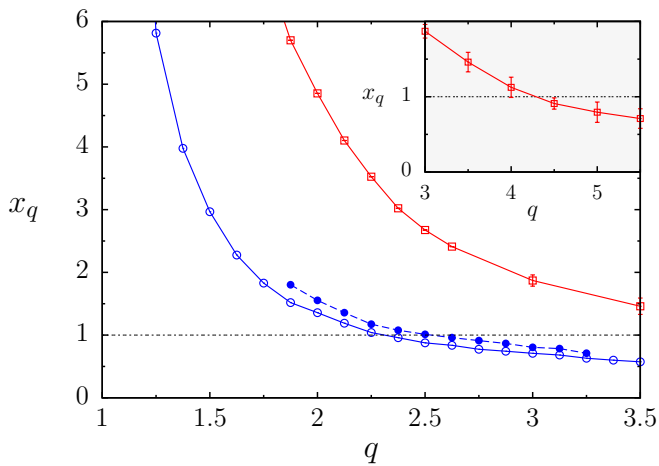}
\end{center}
\caption{(Color online) Tail exponents $x_q$ for eigenvectors of the random
  intermediate map for $\gamma=1/3$ with $N=2^{12}$ (blue filled
  circles) and $N=2^{14}$ (blue empty circles) and for $\gamma=1/11$
  with $N=2^{14}$ (red squares). 
Same parameters as in Fig.~\ref{fig:Dqb}. Inset: Tail exponent for
$\gamma=1/11$ for larger values of $q$. 
\label{fig:xqdirect}}
\end{figure}

\begin{figure}[h]
\begin{center}
\includegraphics[width=\linewidth]{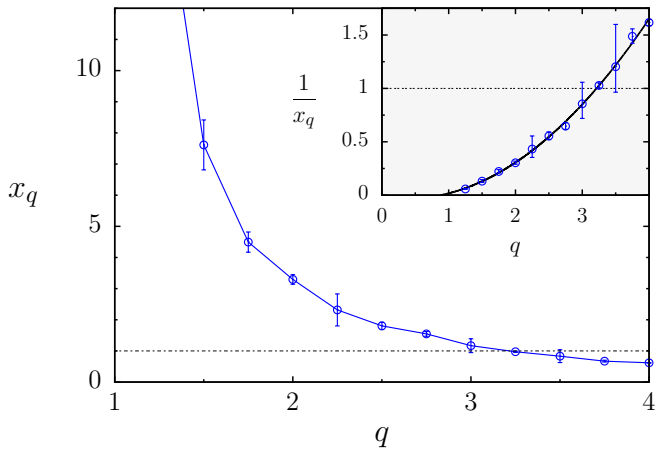}
\end{center}
\caption{(Color online) Tail exponents $x_q$ for iterates of the Anderson map. Same
  parameters as in Fig.~\ref{fig:DqAnd}. The inset shows that $1/x_q$
is well-fitted by a parabola.
\label{fig:xqand}}
\end{figure}

The discrepancy between the two sets of multifractal exponents observed in
the previous section is due to the fact that the moments $P_q$ defined by 
\eqref{moments} have a statistical distribution with a certain
width. In particular for multifractal measures the distribution of the normalized moments
$y_q=P_q/\pqtyp$ is expected to have a power-law tail at large $q$ as
$\mathcal{P}(y_q)\sim1/y_q^{1+x_q}$, with an exponent $x_q$ depending on $q$. 
\begin{figure}[h]
\begin{center}
\includegraphics[width=\linewidth]{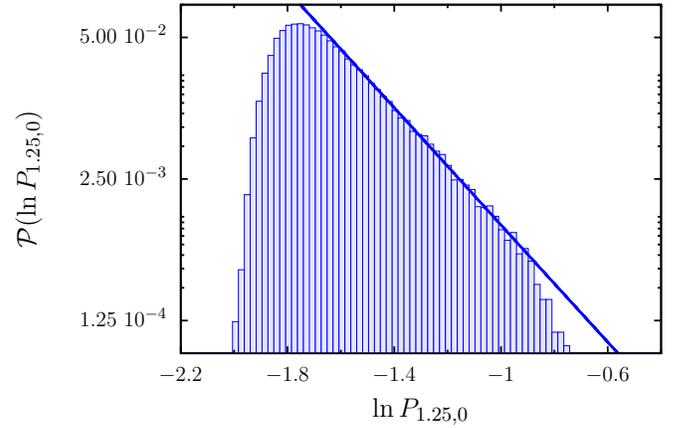}
\end{center}
\caption{(Color online) Probability distribution of the logarithm of the moment $P_q=P(q,0)$
for $q=1.25$ of eigenvectors of the random intermediate map for
$\gamma=1/3$. Same parameters as in Fig.~\ref{fig:Dqb}. The solid line
shows a linear fit (in log scale) whose slope yields the tail exponent
$-x_q$. 
\label{fig:Pmq}}
\end{figure}
In Fig.~\ref{fig:Pmq}
we show an example of the distribution of the logarithm of the moments $P_q$ for the
random intermediate map. The distribution is indeed algebraic with a power law
tail depending on $q$. While the linear behavior (in logarithmic scale) is
clearly observed for small values of $q$ (restricted to the range $q>1$), this is not the case for larger
$q$. We calculated the exponent $x_q$ of the tail for a range of
values of $q$ where this exponent could be extracted. Results are
displayed in  Fig.~\ref{fig:xqdirect} for the intermediate
map and  Fig.~\ref{fig:xqand} for the Anderson map.
The value of $q$ where $x_q \approx 1$ should
correspond to the value $q_{+}$ where the two curves of $D_q$ and $\dqtyp$
(or $\tau_q$ and $\tauqtyp$)
separate. As one can observe in Fig.~\ref{fig:xqdirect} that value of $q$ is rather
difficult to estimate numerically with sufficient accuracy as the exponents $x_q$ did
not converge to a definite value at the largest vector size available
($N=2^{14}$). However the curves seem to yield an exponent equal to 1 around $q\approx
2.51$ for $\gamma=1/3$ and $4.3$ for $\gamma=1/11$. These values are indicated with
black arrows in Fig.~\ref{fig:tauq}, and at that scale they do seem to
coincide with points where $\tau_q$ and $\tauqtyp$ separate. However
these points are far beyond the value $q\approx 1$ at which the multifractal dimensions $D_q$
and $\dqtyp$ separate at the scale of Fig.~\ref{fig:DeltaDq}, and also
different from the values obtained by fixing the relative difference
of the exponents in Fig.~\ref{figdist} (equal to $q_{+ 0.01}=1.31$ for $\gamma=1/3$ and $q_{+ 0.01}=2.51$ for $\gamma=1/11$).
The value where the curves separate is dominated by the rare events in the extreme
tail of the distribution. The different values obtained indicate that indeed the numerical results
are still far from the asymptotic regime. Similar
conclusions can be drawn from  Fig.~\ref{fig:xqand}, which presents the power-law
tail exponents $x_q$ obtained for the moment distribution of the Anderson
vectors: the point where $x_q=1$ is reached around $q\approx 3$ while
Fig.~\ref{fig:DeltaDq} seems to indicate a separation of the
multifractal exponents around $q\approx 1$.  We note that for the finite 
sizes considered, the value of $x_q$ seems to become infinite as 
$q\rightarrow 1$, indicating that in this regime the distribution of moments is not any more 
fitted  by a power law at large moments (see also Fig.~\ref{fig:xq3}).  The
behavior of the exponents $x_q$ will be further discussed in the next Subsection.

\subsection{Relation between multifractal exponents and moment distribution}
As explained in Section \ref{knownfacts}, it was proposed in \cite{MirlinEvers00}
that the exponents $\tau_q$ and $\tau_q^{\rm typ}$ were related to
the moment distribution through the relation (\ref{xqtq}).  This
formula was proved only in some very specific cases, such as the PRBM
model in the regime of weak multifractality, but conjectured
to be generically valid.  The results of the preceding subsections
enable to check numerically whether this formula holds for our models.

\begin{figure}[h]
\begin{center}
\includegraphics[width=8.5cm]{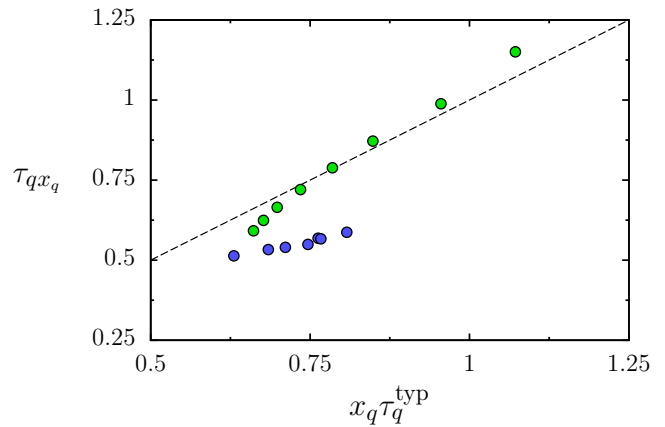}
\end{center}
\caption{(Color online) Relation between $\tau_{qx_q}$ and $x_q\tau_q^{\rm
    typ}$ for eigenvectors of the intermediate map for $\gamma=1/3$. Parameter values
  are the same as in Fig.~\ref{fig:Dqb}. Blue/green (black/grey) circles
  correspond to values of $q$ larger/smaller than $2.5$ (the values closer
to $q=2.5$ are on the left for both series of points). Dashed line is the formula (\ref{xqtq}).\label{fig:xqtq} }
\end{figure}

\begin{figure}[h]
\begin{center}
\includegraphics[width=8.5cm]{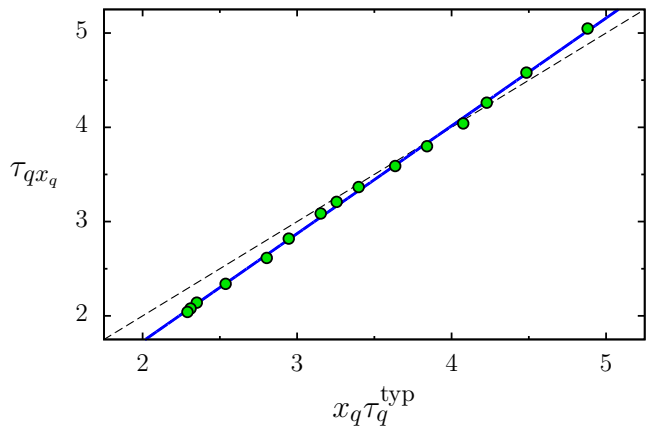}
\end{center}
\caption{(Color online) Relation between $\tau_{qx_q}$ and $x_q\tau_q^{\rm typ}$ for eigenvectors of the intermediate map for $\gamma=1/11$. Parameter values are the same as in Fig.~\ref{fig:Dqb}.
The blue solid line shows a linear fit of the data ($y=1.14x-0.56$). 
Dashed line is the formula (\ref{xqtq}).\label{fig:xqtqb11}}
\end{figure}
In Fig.~\ref{fig:xqtq},\ref{fig:xqtqb11} we show $\tau_{qx_q}$ as a function of $x_q \tauqtyp$ 
for the random intermediate map with parameters $\gamma=1/3$ and
$\gamma=1/11$. For $\gamma=1/3$ (Fig.~\ref{fig:xqtq}), it
shows a certain agreement with the conjectured law for small values of
$q$. Similarly for $\gamma=1/11$ (Fig.~\ref{fig:xqtqb11}) the agreement with the law
\eqref{xqtq} is good, but the actual slope seems slightly different.
On the other hand, for $\gamma=1/3$
and larger values of $q$ the relation breaks down. We note that there is a certain
ambiguity in the formula, since as can be seen in Fig.~\ref{fig:xqtq}
one can have two values of $\tau_{qx_q}$ for the same value of 
$x_q\tau_q^{\rm typ}$ (corresponding to two different values of $q$).  The results indicate that the relation
(\ref{xqtq}) can indeed be seen, even if approximately, in other systems that in Anderson transition
models.  Interestingly, the case $\gamma=1/11$ corresponds to a case of weak
multifractality.
This might indicate that the relation is better verified in the case
of weak multifractality, and only approximate in the general case.
But we cannot exclude that the regime of weak multifractality leads
to weaker finite-size effects and that the results for $\gamma=1/3$ would
eventually converge to the law (\ref{xqtq}) for larger sizes and many more
realizations.  An additional problem concerns the different scales
of the two figures \ref{fig:xqtq} and \ref{fig:xqtqb11}.  
As the multifractality is weaker in the case $\gamma=1/11$, the values of $x_q$ 
and $\tau_q$ are larger, leading to a much larger scale for the data
in Fig.~\ref{fig:xqtqb11}.  It is possible that the finite-size effects
are comparable but show more markedly in Fig.~\ref{fig:xqtq} due to its much smaller scale.

\begin{figure}[h]
\begin{center}
\includegraphics[width=8.5cm]{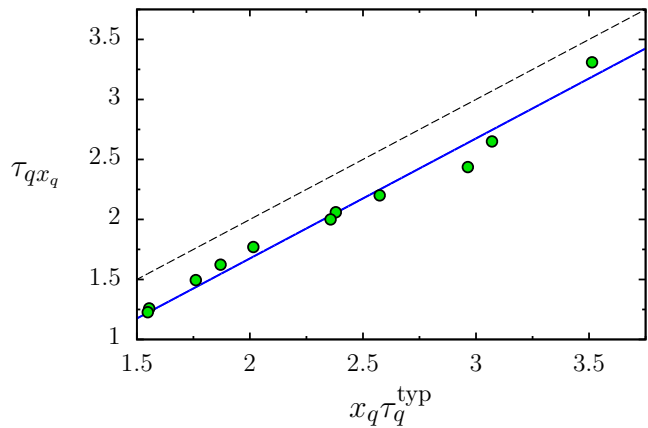}
\end{center}
\caption{(Color online) Relation between $\tau_q$ and $\tau_q^{\rm typ}$ for iterates of the Anderson map with $k=1.81$, $N=2^{11}$, $|\psi(0)\rangle=|i\rangle$ with $i=1024$ and $t=10^8$. Parameter values are the same as in Fig.~\ref{fig:DqAnd}. The blue solid line correspond to $y=x-0.29$. Dashed line is the formula (\ref{xqtq}). \label{fig:xqtqand}}
\end{figure}

In Fig.~\ref{fig:xqtqand} we present the same numerical analysis
for the Anderson map. The results show that a linear law similar to
\eqref{xqtq} can be seen. The slope is close to one, 
but the curve is shifted by a relatively large offset ($\approx 0.3$).  
Note that the formula
(\ref{xqtq}) was predicted for eigenvectors of the Anderson model and
PRBM; here we are looking at iterates of wave packets, 
which can show different behavior.  

\begin{figure}[h]
\begin{center}
\includegraphics[width=\linewidth]{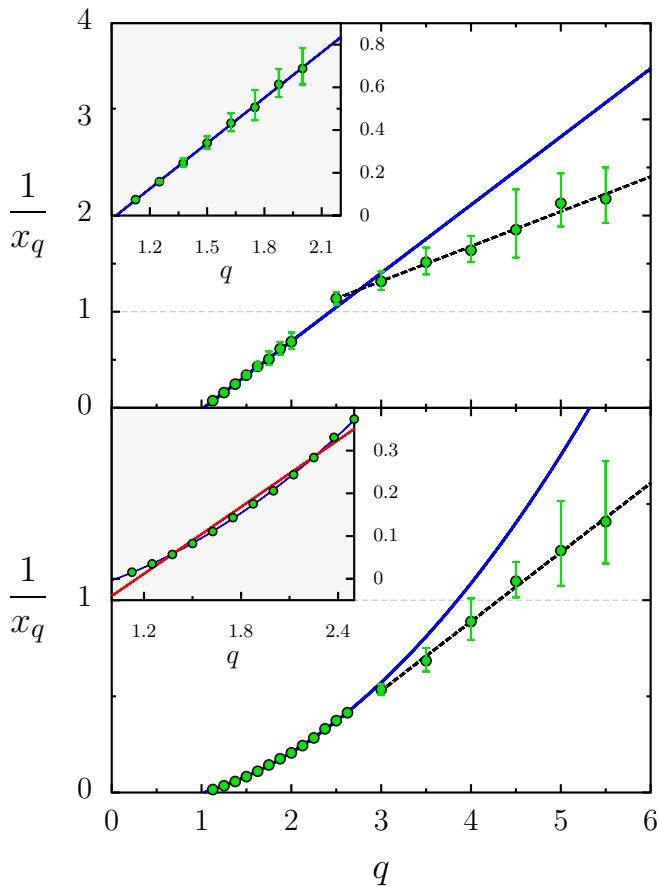}
\end{center}
\caption{(Color online) Inverse of the tail exponents $x_q$ for eigenvectors of the random intermediate map. Top panel: $\gamma=1/3$; blue and black dashed lines are
linear fits in two different $q$ range; dashed line (linear fit for large $q$)
is $1/x_q= 0.36 q +0.23$; inset is a blow-up of the small $q$ range. Bottom panel: $\gamma=1/11$; blue and black dashed lines are
respectively quadratic and linear fits in two different $q$ range;  dashed
line (linear fit for large $q$) is $1/x_q= 0.36 q -0.55$;
inset is a blow-up of the small $q$ range, showing in red the best linear fit for the same data. In both cases, the two ends of the error bars correspond to two values of $x_q$ obtained from a fit of the moments probability distribution over two different intervals. Parameter values are the same as in Fig.~\ref{fig:Dqb}. \label{fig:xq3} }
\end{figure}

Interesting properties of the exponents $x_q$ can be deduced from the
relation \eqref{xqtq}. As mentioned in the introduction, if such a
relation is verified it implies that for $q>q_+$ (or equivalently
$x_q<1$) the inverse of
the exponents $x_q$ should follow a linear law. Indeed, for $q>q_+$
the exponent $\tauqtyp$ is linear and one has 
$qx_q\alpha_{+}=\tau_{qx_q}=D_{qx_q}(qx_q-1)$, thus $z=qx_q$ is solution of
the equation $D_z=\frac{z}{z-1}\alpha_{+}$. If this equation has an
unique solution, then the quantity $qx_q$ is a constant, equal to $q_{+}$ for
$q=q_{+}$, and thus $1/x_q$ should be linear as a function of $q$ for
$q>q_+$.
In order to check whether this holds in our case, we plot 
in Fig.~\ref{fig:xq3} the values of $1/x_q$ as a
function of $q$ for the intermediate random map with parameters $\gamma=1/3$
and $\gamma=1/11$. 
In both cases, a linear law agrees well with the data at large $q$, but 
obeys an equation different from the one predicted, with in particular
an extra constant term which depends on $\gamma$.

A second consequence of relation \eqref{xqtq} arises in the case where
$D_q$ is given by a linear function. This is in particular the case
for the intermediate map in the regime of weak multifractality.
In this regime it was observed \cite{MGG} that for not too large
$q$ the multifractal exponents $D_q$ are very closely given by the linear approximation $D_q=1-q/b$. 
Inserting this relation into \eqref{xqtq}, we get that for
$q\in ]q_-,q_+[$ (where $D_q$ and $\dqtyp$ are equal) the exponents
are given by $x_q=b/q^2$ provided $q x_q$ also belongs to
$]q_-,q_+[$.  We note that this in turn predicts a value
$q_+=\sqrt{b}$ for the separation point between the two sets of exponents, 
contrary to the linear scaling found in the data shown in Fig.~\ref{figdist}.
According to these considerations, in the small $q$ regime and for weak multifractality, a
quadratic behavior of $1/x_q$ should be observed provided the linear
regime extends beyond the point $q_+$.
In the case $\gamma=1/11$ for the intermediate map, the linear regime
is verified quite far away from zero but breaks down before $q_+$ (see
Fig.~\ref{fig:DeltaDq}), thus $D_{q x_q}$ is not a linear function of its argument. Still, Fig.~\ref{fig:xq3} (lower panel) shows that
the parabolic behavior of $1/x_q$ is retained for small values of $q$; 
the inset shows that indeed a quadratic fit is much better than a linear fit
in this range.
In the strong multifractality regime, $D_q$ is not linear any more
beyond $q=1$; in that case, the inset of the top panel in Fig.~\ref{fig:xq3} 
shows that the behavior of $1/x_q$ is linear rather than quadratic.

\subsection{Symmetry between exponents}
As described in Section \ref{knownfacts}, it was predicted analytically
and observed in the Anderson model at the transition that a
symmetry exists between multifractal exponents.  Indeed,
the quantity $\Delta_q=\tau_q-q+1$ follows the law
$\Delta_q=\Delta_{1-q}$ (relation (\ref{deltaqrel})).  This relation was predicted on very
general grounds, and expected to hold for all multifractal quantum systems.
It was seen in the PRBM model \cite{fyodorov}, 
the Anderson model \cite{deltaq} and ultrametric random matrices 
\cite{ossipov}.
It was also predicted to occur in simple multifractal cascade models
in \cite{MonGar10}.
We performed systematic calculation of the quantity $\Delta_q$ for each of
the systems considered. 

\begin{figure}[h]
\begin{center}
        \includegraphics[width=\linewidth]{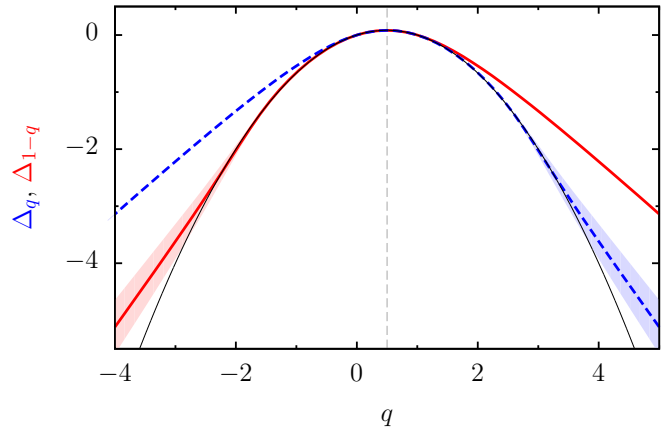}
\end{center}
\caption{(Color online) Anomalous exponents $\Delta_q$ (blue dashed curve) and $\Delta_{1-q}$ (red solid curve) for eigenvectors of the intermediate map with random phases, $\gamma=1/3$ and $N=2^{14}$. Parameter values are the same as in Fig.~\ref{fig:Dqb}. The black thin line shows the parabola $q(1-q)/b$ corresponding to the linear regime.\label{fig:deltaq} }
\end{figure}

Figure \ref{fig:deltaq} shows the results of this analysis for 
the random intermediate map.  The presence of a large linear regime
complicates the picture, since the linear law described above in Subsection
\ref{linear_regime}
verifies the symmetry.  Thus the intermediate map
can show deviations from the symmetry only outside the linear regime.
It turns out (comparing Figs.~\ref{figE1} and \ref{fig:deltaq}) that
the symmetry (\ref{deltaqrel}) is only present in the linear 
regime and does not
extend beyond its validity.
This seems to indicate that the symmetry is absent from these models.

\begin{figure}[h]
\begin{center}
        \includegraphics[width=\linewidth]{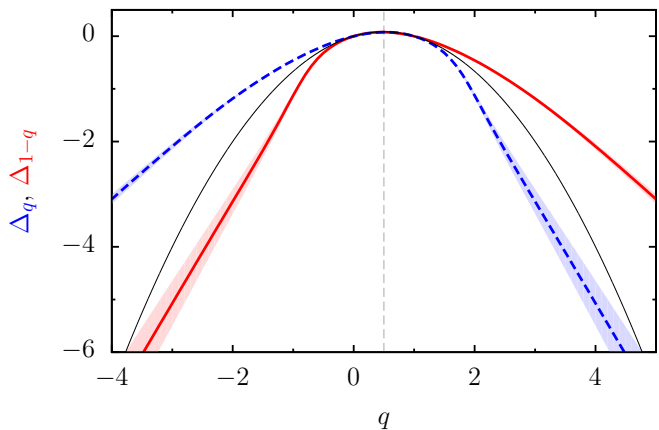}
\end{center}
\caption{(Color online) Anomalous exponents $\Delta_q$ (blue dashed curve) and $\Delta_{1-q}$ (red solid curve) for eigenvectors of the intermediate map (\ref{intermediate_map}) with $\gamma=1/3$ and $N=2^{14}$. The (light blue, light red) shaded region indicates standard error in the least squares fitting. The multifractal analysis was done with box sizes ranging from 16 to 1024. The black thin line shows the parabola $q(1-q)/b$ corresponding to the linear regime.\label{figF2}}
\end{figure}

As the linear regime is much smaller 
in the case of the intermediate map without
random phase, if the symmetry does not hold in this system we should expect
a larger discrepancy in the nonrandom case. Fig.~\ref{figF2} displays
$\Delta_q$ and $\Delta_{1-q}$ for the nonrandom intermediate case, 
showing that indeed the symmetry is verified in an even smaller
range of $q$ values than for the random case, in agreement with
the small extent of the linear regime.

Finally, we have also computed the quantities $\Delta_q$ and 
$\Delta_{1-q}$ for iterates of the random intermediate map and of the 
random Anderson map. As said in Subsection V.A, the precision of the 
exponents degrades for $q \lesssim -2$ in the case of the Anderson map, so
verification of the symmetry relation is delicate. Nevertheless,
our results indicated that while for iterates of the random intermediate map
 the symmetry (\ref{deltaqrel}) still does not hold beyond the linear regime,
in the case of the Anderson map the symmetry remains valid within the numerical error bars (which are however quite large) 
well beyond the linear regime (data not
shown). This seems to indicate further that
the symmetry is a feature of the Anderson model, which is clearly 
absent from
the intermediate map.

Although we cannot prove rigorously that the symmetry relation (\ref{deltaqrel}) does not hold for
intermediate systems, our results strongly indicate that it is violated in these
systems as soon as the linear regime breaks down.  To further confirm that
our numerical method is able to observe the symmetry (\ref{deltaqrel}) in a system where it is present, we
computed $\Delta_q$ and 
$\Delta_{1-q}$ for ultrametric random matrices where the relation is known
to hold \cite{ossipov}.  Our numerical method was able to confirm 
unambiguously the presence
of the symmetry in this specific case (data not shown).

We note that in \cite{fyodorov} the presence of the symmetry (\ref{deltaqrel}) for the
Anderson transition was theoretically predicted on the basis
of a renormalization group flow whose limit corresponds to
a nonlinear sigma model.  It would be interesting to see if
a different nonlinear sigma model can apply to the intermediate map, or
if it is the sign that these models cannot describe certain aspects of these
systems.

\section{Conclusion}

In this paper, we have studied the different multifractal exponents one can extract from the wave functions of the intermediate map and the Anderson
map.  Both models are one-dimensional, and thus allow much larger system size
than the 3D Anderson transition, but in contrast to Random Matrix models such as the PRBM
they correspond to physical systems with an underlying dynamics.

Our results enabled to extract the exponents over a large range of $q$ values
for the intermediate map. We have checked that two methods widely used in other contexts (classical multifractal systems),
namely the box-counting and the wavelet method, can be used to obtain
reliably the exponents, giving similar results in most cases, although
the box-counting method seems more robust for large negative values.   

Our numerical data allow to confirm that the Anderson map introduced
in  \cite{dima} and experimentally implemented with cold atoms \cite{delande}
indeed displays multifractal properties at the transition point, although this
multifractality is weak. As concerns the intermediate map, our data confirm 
the existence of a linear regime for the multifractal exponents $D_q$ and
$\dqtyp$ which was first seen in 
\cite{MGG}, well beyond the regime of weak multifractality.  Interestingly enough, the linear regime is much smaller for the nonrandom intermediate map.
We checked that the exponents $D_q$ and $\dqtyp$ are different also in the case
of the intermediate map, even in the nonrandom case where no disorder is present.  

Our numerical study of the moments of the wave functions and the multifractal
exponents show that the generic behavior of $D_q$ and $\dqtyp$ predicted
for the Anderson transition \cite{mirlinRMP08} are present for the Anderson map.  Our results enable to extract the value of $q_+$ and $q_-$ through the behavior of the moments of the wave functions together with the values of
$\alpha_+$ and $\alpha_{-}$; the fact that the value is different from the one obtained by direct
computation of the multifractal exponents shows that finite size effects persist
in such systems up to very large sizes.  Note, however, that as our 
numerical computations correspond to very large system sizes, this might
indicate that the asymptotic limit may be difficult to reach even in
experimental situations. In addition, our investigations show that the
relation (\ref{xqtq}) between the moments and the exponents 
conjectured in \cite{MirlinEvers00} is only approximately verified in our systems, even in the Anderson map. At last, the exact symmetry relation (\ref{deltaqrel})
between the multifractal exponents of the Anderson transition discovered in \cite{fyodorov} is not present in intermediate systems. 

Our results indicate that intermediate systems, and more generally 
quantum pseudointegrable systems, represent a type of model with some
similarities with the Anderson transition model of condensed matter, but with
specific properties. In particular, the absence of the symmetry present in
the Anderson model between the exponents suggests proceeding with care 
in using the nonlinear sigma model to predict properties of these systems.

We think that further studies of these two kind of
quantum systems are needed in order to elucidate the multifractal 
properties of quantum systems and their link with spectral statistics.

We thank E. Bogomolny, D. Delande, K. Frahm, C. Monthus and A. Ossipov  for useful discussions.
This work was supported in part by the Agence Nationale de la Recherche
(ANR), project QPPRJCCQ and by a PEPS-PTI from the CNRS.


\end{document}